\documentclass [12pt]{article}
\usepackage {amssymb}
\usepackage {amsmath}
\usepackage {graphicx}

\begin{document}

\footnotesize{\noindent V. Krasnoholovets, Reasons for graviational mass and the problem of quantum gravity, 
\textit{Ether, Spacetime and Cosmology}, {\bf Vol. 1}. Eds.: M. C. Duffy, J. Levy and V. Krasnoholovets  
(PD Publications, Liverpool, 2008), pp. 419-450 (ISBN 1 873 694 10 5)}.

\bigskip

\begin{center}
\section*{REASONS FOR GRAVITATIONAL MASS AND
THE PROBLEM OF QUANTUM GRAVITY}
\end{center}

\bigskip

\begin{center}
\textbf{Volodymyr Krasnoholovets}
\end{center}

\bigskip

\begin{center}
Institute for Basic Research \\
90 East Winds Court, Palm Harbor, FL 34683, USA
\end{center}

\bigskip

\begin{flushright}
14 February 2005 \quad\quad \ \ \ \
\end{flushright}

\abstract {The problem of quantum gravity is considered from a
radically new viewpoint based on a detailed mathematical analysis of
what the constitution of physical space is, which has recently been
carried out by Michel Bounias and the author. The approach allows
the introduction of the notion of mass as a local deformation of
space regarded as a tessellation lattice of founding elements
(topological balls, or cells, or superparticles) whose size is
estimated as the value of around the Planck one. The next aspect
used at the examination of the problem is submicroscopic mechanics
developed by the author, which can easily be transformed to the
conventional formalism of quantum mechanics. The interaction of a
moving particle-like deformation with the surrounding lattice of
space involves a fractal decomposition process that supports the
existence and properties of previously postulated ``inerton clouds"
as associated to particles. The cloud of inertons surrounding the
particle spreads out to the range $\Lambda = \lambda \,c/v $ from
the particle where $v$ and $c$ are velocities of the particle and
light, respectively, and $\lambda $ is the de Broglie wavelength of
the particle. Therefore, the particle's inertons return the real
sense to the wave $\psi $-function as the field of inertia of the
moving particle. Since inertons transfer fragments of the particle
mass, they play also the role of carriers of gravitational
properties of the particle. The submicroscopic concept stated above
has successfully been verified experimentally, though so far rather
in microscopic and intermediate ranges.}

\bigskip

\textbf{Key Words:} space, tessel-lattice, quantum gravity,
inertons, submicroscopic mechanics

\bigskip

\textbf{PACS:} 03.65.Bz Foundations, theory of measurement; 04.20.Cv
Fundamental problems and general formalism; 04.60.-m Quantum gravity;

\bigskip

\section*{1. Introduction}

\hspace*{\parindent} Gravity was initially constructed as a
classical theory. However, the discovery of quantum phenomena
allowed researchers to assume the existence of a possible quantum
origin for gravity. This established a new discipline known today
as quantum gravity. Nowadays one can distinguish several major
branches in the study of quantum gravity. Newer approaches abandon
the early idea of general relativity as a dynamical system and
settle up quantum field theories exhibiting general covariance
and/or manifolds. Concepts developed in the particle physics,
which investigate possibilities of propagation of quantum fields
with the specification of their internal structure, such as gauge
group, fields, and interactions, attract a wider interest.
Significant role places symmetries of quantum field theories,
which reflect the mappings of manifolds. Thus quantum gravity is
related to manifold topology. Good reviews on modern approaches to
the description of quantum gravity, such as quantization of
gravity, the non-renormalisability of gravity, supergravity,
superstring theory (or M-theory), loop quantum gravity, gauge
theories and others, one can find, e.g. in Refs. [1-4]; a review
of latest trends in the study of quantum space-time is available
in Balachandran [5].

Sarfatti [6,7] has sketched very recent progress in the foundations of
quantum theory representing his views on Einstein's gravity including both
dark energy and dark matter, which emergent macro-quantum phenomena in the
sense of Penrose's ``off diagonal long range order" inside the vacuum and
this is in line with Anderson's ``more is different" and Sakharov's ``metric
elasticity" approaches. His result is a background-independent
non-perturbative model that agrees loop quantum gravity. He offers a model
for the formation of the vacuum condensate ``inflation scalar field" filled
with bound virtual electron-positron pairs; these pairs generate the
globally flat quantum electro-dynamic vacuum unstable. By Sarfatti,
Einstein's c-number space-time emerges as the more stable vacuum (quite
similar to the superconducting vacuum state for a condensate of real
electron pairs where the attractive virtual phonons overpower the repulsive
virtual photons in the real electron-electron pair coupling). Sarfatti's
means are de Broglie-Bohm's pilot wave interpretation of quantum theory and
the usual ``two fluid model" in which the residual random zero-point vacuum
fluctuations of all available dynamical fields are the ``normal fluid". Thus
Sarfatti seems reasonable concludes: ``There is no such thing as a local
classical world. We live in a local macro-quantum world!" His results [6,7]
are close to Kleinert's and Zaanen's [8] who have developed the 4D ``world
liquid crystal lattice" model. This model of Einstein's gravity is
consistent with a foam loop quantum gravity. By Kleinert and Zaanen, general
relativity in which the perfect world crystal lattice with unit cells at the
Planck scale and no topological defect string vortex lines corresponds to
globally flat space-time with zero curvature in all components not only
space curvature. It is interesting that in this model, Einstein's curvature
corresponds to a stringy disclination defect density in the long wave limit
of the elastic world crystal distortion.

Thus gravity is still considered different from the other physical
forces, whose classical description involves fields (e.g. electric
or magnetic fields) and field carriers propagating in space-time.
General relativity says that the gravitational force is related to
the curvature of space-time itself, i.e. to its geometry. We can
see that in gravitational physics space-time is treated as a
dynamical field, though in non-gravitational physics space-time
emerges as the scene on which all physical processes take place.
Therefore available approaches to quantum gravity represent a
collection of liable geometries of space-time. It is pity that
such interesting models of space-time and quantum gravity seems
remain metaphysical by its nature having no contact with
experiments in principle.

Toh [3] said right after C. J. Isham that rather than quantizing
gravity, one should seek a quantum theory that would yield general
relativity as its classical limit. He then adds that the main
obstruction here is the lack of a starting point to construct such
a quantum theory. A radically new theory may require the
re-examination, from their foundation, the concepts of space,
time, and matter [9].

In the present paper we give a demonstration of an alternative viewpoint on
quantum gravity, in fact a tenable starting point for a new quantum theory,
which, in contrast to conventional approaches, allows direct test
experimentation. Experimental aspects and the verification of the theory
presented receive a detailed consideration.

\section*{2. Preliminaries}

\subsection*{2.1. Constitution of physical space and
the foundations of space-time}

\hspace*{\parindent} Physical mathematics developed in the
20$^{th}$ century is still treated as effective means at
theoretical studies in particle and gravitational physics. Various
ill-determined fields, such as the wave $\psi$-function and spinor
$\bar {\psi} $ and more abstract, their gauge transformations,
abstract entities (strings, loops, and etc.), varied kinds of
symmetries and so on, are those host elements of modern
theoretical constructions, which nowadays are used at mathematical
modelling of space-time and fundamental interactions.

Having developed an alternative concept of quantum gravity, we have to
elaborate a new approach to space-time. Starting from pure mathematics,
namely, set theory, topology and fractal geometry, in which any element is
strictly defined, Bounias and the author [10-13] have shown that a physical
space can exist as a collection of closed topologies in the intersections of
abstract topological subspaces provided with non-equal dimensions.

It seems that Nottale [14] was the first who applied fractal geometry to the
description of space-time. He studied relativity in terms of fractality
basing his research on the Mandelbrot's concept of fractal geometry. Nottale
introduced a scale-relativity formalism, which allowed him to propose a
special quantization of the universe. In his theory, scale-relativity is
derived from a special application of fractals. In his approach, the fractal
dimension \textit{D} is defined from the variation with resolution of the
main fractal variable, i.e. the length \textit{l} of fractal curve plays a
role of a fractal curvilinear coordinate. Nottale's approach leads to the
conclusion that a trajectory of any physical system diverges due to the
inner stochastic nature that is caused by the fractal laws. Again, in this
approach fractality is associated with the length of a curve as such.

In our works [10-13] we have shown that fractal geometry can be derived from
complete other mathematical principles, which becomes possible on the basis
of reconsidered fundamental notions of space, measure, and length; this
allowed us to introduce deeper first principles for the foundations of
fractal geometry.

First of all we argue [10-13] that since in mathematics some set
does exist, a weaker form can be reduced to the existence of the
empty set $\O$. An abstract lattice of empty set cells $\O$ has been
shown to be able to account for a primary substrate in a physical
space. Providing the empty set with a special combination rules
including the property of complementary allowed us to define a
magma. The magma ${\O}^{\O} = \{ \complement,{\O}\} $ constructed
with the empty hyperset together with the axiom of availability form
a fractal tessellation lattice. Space-time is represented by ordered
sequences of topologically closed Poincar\`{e} sections of this
primary space. We have demonstrated that the antifounded properties
of the empty set provide existence to a lattice involving a
tessellation of the corresponding abstract space with empty
topological balls. The founding element, i.e. a topological ball, or
a primary cell, of this mathematical lattice called the
\textit{tessel-lattice}, can be estimated at the size close to the
Planck one, or more exactly, $\sim $ 10$^{-35}$ m. Discrete
properties of the tessel-lattice allow the prediction of scales at
which microscopic to cosmic structures should occur.

Deformations of primary cells by exchange of empty set cells allow a
cell to be mapped into an image cell in the next section as far as
mapped cells remain homeomorphic. However, if a deformation involves
a fractal transformation to objects, there occurs a change in the
dimension of the cell and the homeomorphism is not conserved. Then
the fractal kernel stands for a \textit{particle} and the reduction
of its volume (together with an increase of its area up to infinity)
is compensated by morphic changes of a finite number of surrounding
cells. Therefore, the state described is associated with the
availability of an actual \textit{mass} in the tessel-lattice, i.e.
a local deformation of the tessel-lattice (a deformation of a cell)
is identified with what we call ``mass" in conventional physics. The
interaction of a moving particle-like deformation with the
surrounding lattice involves a fractal decomposition process that
supports the existence and properties of previously introduced
\textit{inerton clouds} [15,16] as associated to particles.

Thus the tessel-lattice is found in the degenerate state that does
not manifest itself as matter; cells of the tessel-lattice are empty
sets and this is a typical vacuum with its flat geometry. However,
the tessel-lattice emerges when a deformed cell (i.e. a massive
particle) starts to move: the decomposition of this local moving
deformation entrains other cells of the tessel-lattice. Such a
decomposition, or in the physics language clouds of inertons, should
allow the experimental verification. We will back to this issue
below.

The existence of closed topological structures gives the intersection of two
spaces having nonequal dimensions owns its accumulation points and is
therefore closed. In other words, the intersection of two connected spaces
with nonequal dimensions is topologically closed. This allowed the
representation of the fundamental metrics of our space-time by a convolution
product where the embedding part
\begin{equation}
\label{eq1} D4 = \int {{\kern 1pt} \left( {\int\limits_{{\kern 3pt}
d{\kern 1pt} V} {dx{\kern 1pt} {\kern 1pt} dy{\kern 1pt} {\kern 1pt}
dz{\kern 1pt} {\kern 1pt}} }  \right)} \ast d\psi \left( {w} \right)
\end{equation}

\noindent where \textit{d{\kern 0.6pt}V} is the element of
space-time and $d\psi \left( {w} \right)$ the function accounting
for the extension of 3-$D$ coordinates to the 4$^{\rm th}$ dimension
through convolution ($ \ast $) with the volume of space.

Time is an emergent parameter indexed on non-linear topological
structures guaranteed by discrete sets. This means that the
foundation of the concept of time is the existence of order
relations in the sets of functions available in intersect
sections. The symmetric difference between sets and its norm can
be treated as a new, more general, \textit{non-metric distance}.
As the set distance \textit{D} is the complementary of objects,
the system stands as a manifold of open and closed subparts.
Mapping of these manifolds from one to another section, which
preserve the topology, represents a reference frame in which
topology has allowed one to specify the changes in the
configuration of main components: if morphisms are observed, then
this enables the interpretation as a motion-like phenomenon, when
one compares the state of a section with the state of mapped
section. Hence a space-time-like sequence of Poincar\'{e} sections
is a non-linear convolution of morphisms. This means that time is
not a primary parameter. And the physical universe has no more
beginning: time is just related to ordered perceptions of
existence, not to existence itself. The topological space does not
require any fundamental difference between reversible and steady
state phenomena, nor between reversible and irreversible process.
Rather relation orders simply hold on non-linearity distributed
topologies, and from rough to finest topologies.

The mapping of two Poincar\'{e} sections is assessed by using a
natural metrics of topological spaces. Let $\Delta \rm \left(
{A,B,C,...} \right)$ be\textbf{\textit{} }the generalized set
distance as the extended symmetric difference of a family of
closed spaces,
\begin{equation}
\Delta({\rm A}_{i})_{{\kern 1pt}i {\kern 1pt} \in {\kern 1pt} {\rm
N}}={\mathop \complement_{\cup \{ {\rm A}i \}}}{\kern 1pt}
{\mathop \cup _{{\kern 2pt} i\neq j}}{\kern 1pt}({\rm A}_i \cap
{\rm A}_j). \label{2}
\end{equation}

The complementary of $\Delta $ in a closed space is closed. It is
also closed even if it involves open components with nonequal
dimensions. In this system ${\mathfrak{m}}\langle \left( {{\rm
A}_{i}} \right)\rangle = \Delta $ has been associated with the
instant, i.e., the state of objects in a timeless Poincar\'{e}
section. Since distances $\Delta $ are the complementaries of
objects, the system stands as a manifold of open and closed
subparts.

The set-distance provides a set with the finer topology and the set-distance
of nonidentical parts provides a set with an ultrafilter. Regarding a
topology or a filter founded on any additional property $\left( { \bot
{\kern 1pt} {\kern 1pt} {\kern 1pt}}  \right)$ this property is not
necessarily provided to a $\Delta $-filter. The topology and filter induced
by $\Delta $ are thus respectively the finer topology and an ultrafilter.
The mappings of both distances and instants from one to another section can
be described by a function called the \textit{moment of junction}, since it
has the global structure of a momentum [11]. The moment of junction allowed
us to investigate the composition of indicative functions of the position of
points within the topological structures and to account for elements of the
differential geometry of space-time. Therefore, the moment of junction
enables the formalization of the topological characteristics of what is
called \textit{motion} in a physical universe. It is the motion that has
been called by de Broglie as the major characteristic determining physics.
While an identity mapping denotes an absence of motion, that is a null
interval of time, a nonempty moment of junction stands for the minimal of
any time interval. Sidharth [17] tried to argue that a minimum space-time
interval should exist and that ``one cannot go to arbitrarily small
spacetime intervals or points''. In our sense, there is no such ``point'' at
all: only instants [10,11] that at bottom of fact do not reflect timely
features.

Space-time versus neo-Lorentzian interpretation of relativity is conciliated
in the tessellation space. Indeed, according to Bouligand, Minkowski,
Hausdorff and Besicovich when all intervals (at least nearly) have the same
size, then the various dimension approaches are reflected in the resulting
relation [11]
\begin{equation}
\label{eq3} n{\kern 1pt} {\kern 1pt} \cdot \left( {\rho}
\right){\kern 1pt} ^{\rm e}\, \cong 1.
\end{equation}

Here e is the Bouligand similarity (integer) exponent that
specifies a topological dimension and this means that a
fundamental space E can be tessellated with an entire number of
identical topological balls exhibiting a similarity with E, upon
the similarity coefficient $\rho $; $n$ is the number of parts.
The fractality of particle-giving deformations gathers its spatial
parameters and velocities into a self-similarity expression (on
the basis of relation (3)), which provides the space-to-time
connection required by special relativity.

Indeed, the fractality of particle-giving deformations gathers its
spatial parameters $\varphi _{{\kern 0.5pt} i} $ and velocities
$\upsilon $ into a self-similarity expression that provides a
space-to-time connection. Indeed, let $\varphi_{{{\kern 0.5pt}0}}$
and $v_{{\kern 0.5pt}0}$ be the reference values; then the
similarity ratios are $\rho  \left( {{\kern 1pt} \varphi } \right)
= \left( {\varphi _{{\kern 1pt} i} /\varphi_ {\kern 0.5pt 0}}
\right)$ and $\rho \left( {{\kern 1pt} v {\kern 1pt}} \right) =
\left( {v/v _{\kern 0.5pt 0}} \right)$. Therefore,
\begin{equation}
\label{eq4} \rho {\kern 1pt} \left( {{\kern 1pt} \varphi {\kern
1pt}} \right)^{{\kern 0.6pt} \rm e} + \rho {\kern 1pt} \left(
{{\kern 1pt} v {\kern 1pt}} \right)^{{\kern 0.6pt} \rm e} = 1.
\end{equation}

The left-hand side of Eq. (4) includes only space and space-time
parameters, because $\left( {{\kern 1pt} \varphi {\kern 1pt} _{i}}
\right) = $\{distances (\textit{l}) and masses (\textit{m})\}.
Hence we can set $m_{{\kern 1pt}0} /{\kern 1pt} m\,\,\, =
\,\,\,l/{\kern 1pt} l_{{\kern 1pt} 0} $, so that
\begin{equation}
\label{eq5} {\kern 1pt} \left( {m_{0} /m{\kern 1pt}} \right)^{\,
\rm e} + {\kern 1pt} {\kern 1pt} \left( {v /{\kern 1pt} v _{{\kern
0.6pt}0} {\kern 1pt}} \right)^{{\kern 1pt}\rm  e} = {\kern 1pt}
\left( {l_{0} /{\kern 1pt}l{\kern 1pt}} \right)^{\, \rm e} +
{\kern 1pt} {\kern 1pt} \left( {v /{\kern 1pt} v_{{\kern 1pt}0}
{\kern 1pt}} \right)^{{\kern 1pt}\rm  e} = 1.
\end{equation}

The geometry at $\rm e > 1$ escapes the usual space-time; the
necessity of an embedding 4-D timeless space (1) ensures the
coefficient e to be equal to 2 (see Ref. [11] for details). Then
the boundary conditions give rise to the following results for
distances $l,\,\,l_{0} $ and masses $m,\,\,\,m_{{\kern 1pt} 0}$:
\begin{equation}
\label{eq6} {\kern 1pt} \left( {m_{0} /m{\kern 1pt}}  \right)^{2}
+ {\kern 1pt} {\kern 1pt} \left( {\upsilon /\upsilon _{0} {\kern
1pt}}  \right)^{2} = 1\quad \Leftrightarrow {\kern 1pt} \quad m =
m_{{\kern 1pt} 0} /\sqrt {1 - \left( {\upsilon /\upsilon _{0}
{\kern 1pt}}  \right)^{2}} ;
\end{equation}
\begin{equation}
\label{eq7} {\kern 1pt} \left( {l/l_{0} {\kern 1pt}}  \right)^{2}
+ {\kern 1pt} {\kern 1pt} \left( {\upsilon /\upsilon _{0} {\kern
1pt}}  \right)^{2} = 1\quad \Leftrightarrow {\kern 1pt} \quad l =
l_{{\kern 1pt} 0} \sqrt {1 - \left( {\upsilon /\upsilon _{0}
{\kern 1pt}}  \right)^{2}} .
\end{equation}

The Lagrangian ${\kern 1pt} {\kern 1pt} L{\kern 1pt} {\kern 1pt} $
should obey a similar ratio (\ref{eq5}), such that
\begin{equation}
\label{eq8} \left( L{\kern 1pt} /{\kern 0.6pt} L_{{\kern 1pt} 0}
{\kern 1pt} \right)^{{\kern 1pt} 2} + \left( {v /{\kern 0.6pt}v
_{\kern 1pt 0} {\kern 1pt}} \right)^{{\kern 1pt} 2} = 1.
\end{equation}

If we take $L_{{\kern 1pt} 0}  = -  m {\kern 1pt} v_{{\kern
0.6pt}0}^{{\kern 0.6pt}2} $, we obtain
\begin{equation}
\label{eq9} L_{{\kern 1pt} 0}  = - {\kern 1pt} m {\kern 1pt} v
_{{\kern 0.6pt}0}^{{\kern 0.6pt}2} {\kern 1pt} \sqrt { 1 - \left(
{v /v _{{\kern 1pt} 0} } \right)^{2}} .
\end{equation}

By analogy with special relativity, $m$ $v$ and $L$ are the
parameters of a moving object and $ v_{{\kern 1pt} 0}$ is the
speed of light.

To summarise, topology, set theory and fractal geometry allow the
construction of the new theory of physical space [10-13], which in fact
emerges as a mathematical lattice:
\begin{equation}
\label{eq10} \rm F\left( {U} \right) \,\, \cup \,\, \left( {W}
\right)\,\,\, \cup \,\,\left( {c} \right)
\end{equation}

\noindent
where (c) is the set with neither members nor parts, accounts for
relativistic space and quantic void, because (i) the concept of distance and
the concept of time have been defined on it and (ii) this space holds for a
quantum void since it provides a discrete topology, with quantum scales and
it contains no "solid" object that would stand for a given provision of
physical matter. The sequence of mappings of one into another structure of
reference (e.g. elementary cells) represents an oscillation of any cell
volume along the arrow of physical time. This rigorously supports
submicroscopic (quantum) mechanics constructed by the authors in a series of
recent works [15,16,18-20].

\subsection*{2.2. Principle of equivalence}

\hspace*{\parindent} Very recently de Haas [21] studying paradoxes
in the early writings (the first third of the 20$^{\rm th}$ century)
on the relativity and quantum physics has revealed that a
combination of the Gustav Mie's [22] theory of gravity and the Louis
de Broglie's [23] harmony of phases of a moving particle results in
the principle of equivalence for quantum gravity. De Haas notes that
Mie's finding in the description of gravity has been very important
and been used by Gilbert to transform the Mie's Hamiltonian
variation principle into a general covariant variation principle,
which then has been assimilated by other physicists. De Haas
considering Mie's variation principle has shown that the Mie-de
Broglie version of the Hamilton variation principle directly hits in
the quantum domain, that is
\begin{equation}
\label{eq11} \delta \int {H{\kern 1pt} d{\kern 0.5pt}\tau \,
\propto \,{\kern 1pt} } \hbar ,
\end{equation}

\noindent though in the classical limit $\hbar \to 0$ one gets a
conventional non-quantum version of the Hamiltonian principle,
$\delta \int {H{\kern 1pt} d{\kern 0.5pt}\tau } = 0$. Thus the
Mie-de Broglie theory of quantum gravity analyzed by de Haas
accounts for a principle of equivalence of gravitational $m_{\rm
gr}$ and inertial m$_{{\kern 0.5pt}\rm in}$ masses. Namely, the
equality $m_{{\kern 0.5pt}\rm gr} = m_{{\kern 0.5pt}\rm in} $,
which is held in a rest-frame of the particle in question, becomes
invalid in a moving reference frame. In the quantum context, this
equality should be transformed to the principle of equivalence of
the appropriate phases, $\varphi _{{\kern 0.5pt} \rm gr} = \varphi
_{{\kern 0.5pt} \rm in}$. This relation ties up the gravitational
and inertial energies of the particle and also shows that the
gravitational mass is complete allocated in the inertial wave that
guides the particle. Mie's original result for the gravitational
mass was (see sections 42 to 45 in Ref. [22])
\begin{equation}
\label{eq12} m_{{\kern 0.5pt}\rm gr} = m_{0} \sqrt {1 -
v^{2}/c^{2}} ,
\end{equation}

\noindent though the inertial mass was identified with the total
mass, $m_{\rm in} \break = m_{0} /\sqrt {1 - v ^{2}/c^{2}} $. Thus
the gravitation as such seems in fact is a pure dynamic
phenomenon, which on the quantum level has to be associated with
the matter waves.

Consequently, de Haas [21] has revived de Broglie's initial
physical interpretation of wave/quantum mechanics, i.e. the pilot
wave interpretation of de Broglie, and also introduced a fresh
wind in the problem of quantum gravity. By de Haas, it turns out
that a particle moving through the space deforms the metrics on a
quantum local scale in a way that the inertial energy flow
$E_{{\kern 0.5pt} \rm in} {\vec v}_{\rm group} $ becomes
concentrated in the particle's wave packet, though the
gravitational energy flow $E_{{\kern 0.5pt}\rm  gr} {\vec v}_{\rm
particle} $ grows dislocated in it. Nevertheless, on a macroscopic
scale, the equality between these two kinds of energy is preserved
of course. The result of de Haas sounds in fact very
realistically, because he could easy to link Mie's source of
gravitational energy to the trace of inertial stress-energy tensor
that places the role of the source of gravitational energy in
modern concepts.

It is interesting to note that de Haas' finding, i.e. the principle
of equivalence of appropriate phases of inertial and gravitational
energies, on both ontological and mathematical levels strongly
supports the author's own researches [15,16,18-20] on submicroscopic
mechanics of particles moving in the tessellation space. For the
major part, this is the interpretation of the matter waves as a
complicated system that consists of the particle and its inerton
cloud, which move in the space tessel-lattice periodically
exchanging energy and momentum due to the strong interaction with
the tessel-lattice's cells.

\section*{3. The notion of mass}

\hspace*{\parindent} A particulate ball (i.e. a deformed cell of the
tessel-lattice), as indicated above [10-13], provides formalism
describing the elementary particles. In this respect, mass is
represented by a fractal reduction of volume of a topological ball
(or, in other words, a cell of the tessel-lattice), while just a
reduction of volume as in degenerate cells is not sufficient to
provide mass, because a dimensional increase is a necessary
condition. The mass $m_{A} $ of a particulate ball A is a function
of the fractal-related decrease of the volume of the ball,
\begin{equation}
\label{eq13} m_{A} \propto \,{\kern 1pt} {\kern 1pt}
\,\frac{{1}}{{{V}_{\rm particle}} }{\kern 1pt} {\kern 1pt} {\kern
1pt} {\kern 1pt} {\kern 1pt} \left( {{\kern 1pt} {\rm e}_{\kappa}
- 1} \right){\kern 1pt} {\kern 1pt} _{{\rm e}{\kern 0.4pt}
_{\kappa} {\kern 1pt} \ge {\kern 1pt} {\kern 1pt} {\kern 1pt} 1}
\end{equation}

\noindent where e is the Bouligand exponent, and $\left( {{\rm
e}_{\kappa}  - 1} \right)$ the gain in dimensionality given by the
fractal iteration; the index $\kappa $ denotes the possible fractal
concavities affecting the particulate ball.

In condensed matter physics, any foreign particle introduced in the
lattice leads to its deformation that extends from several to many
lattice constants. Such a deformation is called the
\textit{deformation coat} in solids or the solvation in liquids.
This takes place in an ordered/disordered lattice and the network of
molecules. Therefore one can treat such a deformation as a typical
one for any kind of a lattice and seems there are no reasons to
except the tessel-lattice from this rule.

In that way a particulate cell in the tessel-lattice has to generate
its own deformation coat; it has been argued [15,16,18-20] that the
radius of the coat coincides with the Compton wavelength $\lambda
{\kern 1pt} _{\rm Com} = h/m{\kern 1pt} c$ of the particle under
consideration. In the deformation coat all cells are found in the
massive state, though beyond the coat the tessel-lattice's cells are
still massless. That is why this deformation coat can be called a
\textit{crystallite} in which its entities (i.e. massive cells)
undergo oscillations, as is the case with the conventional crystal
lattice. The total mass of the crystallite's cells is equal to the
mass of the particle [18].

A moving particle drags the crystallite all over the place: The
relay readjustment of cells from the massless to massive state and
again to the massless one occurs along the velocity vector
\textbf{v} of the particle. In transversal directions the state of
superparticles remains practically unaltered: cells surrounding
the particle continuously save the same massive state and hence in
these directions cells are rather hard. That is why it is
reasonable to assume that cells in the crystallite vibrate in
transversal vibrations. These transversal vibrations of the
crystallite can be called the (transversal) oscillating mode [24].
The energy of the oscillating mode of the crystallite is $E =
\hbar {\kern 1pt} \omega $ where $E = m_{{\kern 1pt} 0} {\kern
1pt} c^{{\kern 0.6pt}2}/\sqrt {1 - v^{2}/c^{{\kern 0.6pt}2}} $ is
the total energy of the moving particle; in the immobile state $E
\to E_{{\kern 0.6pt} 0} = \hbar {\kern 1pt} \omega _{{\kern 1pt}
0} = m_{{\kern 0.5pt} 0} {\kern 0.5pt} c^{{\kern 0.6pt}2}$[18].

\section*{4. Motion of a particle and the field of inertia}

\hspace*{\parindent} The motion of a particle must involve the
interaction with the ambient space of course. A detailed study of
this motion, which results in submicroscopic mechanics, is described
in Refs. [15,16,18-20]. The appropriate Lagrangian, which contains
terms describing the particle, its interaction with the
tessel-lattice and the cloud of spatial excitations generated at
collisions of the moving particle with the tessel-lattice, can be
written in the form [16]
\begin{equation}
\label{eq14} L = - m_{0} {\kern 1pt} c^{2}\left\{ {1 -
\frac{{1}}{{m_{0} c^{2}}}\left[ {m_{0} \dot {x}^{2} - \frac{{2\pi}
}{{T}}\sqrt {m_{0} \mu _{0}}  {\kern 1pt} \left( {x\dot {\chi}  +
v {\kern 1pt} \chi}  \right) + \mu _{0} \dot {\chi }^{2}} \right]}
\right\}^{1/2};
\end{equation}

\noindent here $m_{{\kern 1pt} 0} ,$ \textit{x} and $\dot {x}$ are
the mass, the position and the velocity of the particle at a
moment \textit{t}, where time \textit{t} is considered as a
natural parameter; $\mu _{{\kern 1pt} 0} ,$ $\chi $ and $\dot
{\chi} $ are the mass, the position and the velocity of
centre-of-mass of the cloud of spatial excitations called
\textit{inertons}; \textit{T} is the period of collisions between
the particle and its inerton cloud; $\upsilon $ is the initial
velocity of the particle; \textit{c} is the velocity of inertons,
which may exceeds the speed of light. The value of the Lagrangian
(14) coincides with that given formally by expression (9).

The Euler-Lagrange equations constructed for variables $q,\,\,\dot {q} = \{
x,{\kern 1pt} {\kern 1pt} {\kern 1pt} \dot {x}{\kern 1pt} ;\,\,\chi
,\,\,\dot {\chi} \} ,$ i.e.
\begin{equation}
\label{eq15} \frac{{d}}{{d{\kern 1pt} t}}\frac{{\partial
L}}{{\partial {\kern 1pt} \dot {q}}} - \frac{{\partial
L}}{{\partial {\kern 1pt} q}} = 0
\end{equation}

\noindent
yield the following uninterrupted solutions
\begin{equation}
\label{eq16} x\left( {t} \right) = v {\kern 1pt} t + \frac{{v
{\kern 1pt} T}}{{\pi} }\left\{ {\left( { - 1} \right)^{{\kern 1pt}
\left[ {{\kern 1pt} t/T} \right]}\cos\left( {\pi {\kern 1pt} t/T}
\right) - \left( {1 + 2{\kern 1pt} {\kern 1pt} \left[ {t/T}
\right]} \right)} \right\},
\end{equation}
\begin{equation}
\label{eq17} \dot {x}\left( {t} \right) = v \cdot \left( {1 -
\left| {{\kern 1pt} {\kern 1pt} \sin\left( {\pi {\kern 1pt} t/T}
\right){\kern 1pt}}  \right|} \right),
\end{equation}
\begin{equation}
\label{eq18} \chi \left( {t} \right) = \frac{{\Lambda} }{{\pi}
}{\kern 1pt} {\kern 1pt} {\kern 1pt} \left| {{\kern 1pt} {\kern
1pt} \sin\left( {\pi {\kern 1pt} t/T} \right){\kern 1pt}}
\right|,
\end{equation}
\begin{equation}
\label{eq19} \dot {\chi} \left( {t} \right) = \left( { - 1}
\right)^{\left[ {{\kern 1pt} t/T} \right]}{\kern 1pt} {\kern 1pt}
{\kern 1pt} c {\kern 1pt} \cos\left( {\pi {\kern 1pt} t/T} \right)
\end{equation}

\noindent where the notation ${\kern 1pt} {\kern 1pt} \left[ {t/T}
\right]$ means an integral part of the integer $t/T.$ Besides, in
expressions (16) to (19),
\begin{equation}
v {\kern 1pt} T = \lambda  \quad {\rm and} \quad c{\kern 0.6pt}T =
\Lambda . \label{20}
\end{equation}

It is readily seen that parameters $\lambda $ and $\Lambda $ play
the role of free pass lengths of the particle and the inerton
cloud, respectively. The solutions (16) to (19) show that the
moving particle periodically exchanges the velocity (and hence the
kinetic energy and the momentum) with its inerton cloud. That is,
the particle emits inertons within each odd sections $\lambda /2$
of the particle path and its velocity gradually decreases from
$\upsilon $ to 0; then the particle absorbs inertons within even
sections $\lambda /2$ of its path and the particle velocity
progressively increases from 0 to $\upsilon $. Besides, the
following relationship comes out from expressions (20),
\begin{equation}
\label{eq21} \Lambda = \lambda {\kern 1pt} {\kern 1pt} c/{\kern
1pt}v,
\end{equation}

\noindent
which connects the particle's and the inerton cloud's free path lengths.

Orthodox quantum mechanics emerges from the canonical Hamiltonian
obtained from the Lagrangian (14) if we pass to the
Hamilton-Jacobi formalism [15,16,18]. Such a transition allows us
to derive de Broglie's relationships
\begin{equation}
E = h\nu \quad {\rm and} \quad   \lambda = h/m v. \label{22}
\end{equation}

Thus submicroscopic mechanics shows that the de Broglie wavelength
is nothing else as the free path length, or the spatial amplitude,
$\lambda $ of the particle. The free path length of the inerton
cloud, or the amplitude of the inerton cloud, $\Lambda $ (21)
characterizes the distance to which the particle's inertons spread
from the particle. \textit{E} from de Broglie's relationships (22)
has been determined as $m\upsilon ^{2}/2$, the mass \textit{m} is
the total mass, which is also the inertial mass, $m = m_{{\kern 1pt}
0} /\sqrt {1 - v^{2}/c^{{\kern 0.6pt}2}}$, that agrees with Mie's
result [22]; $\nu $ is linked with the period of collisions
\textit{T}, namely, $\nu = 1/2{\kern 0.6pt}T$. Planck's constant
\textit{h} gets an interpretation of the minimum increment of the
particle action in the cyclic process, i.e. the motion with the
periodic exchange of energy between the particle and the accompanied
cloud of inertons, which is only guided by the space tessel-lattice.

The availability of relationships (22) allows one immediately to
derive the Schrodinger equation [25], Lorentz-invariant in our
case due to the invariant time entered the equation. Then the wave
$\psi$-function, which so far has treated as pure abstract,
receives the rigorous physical interpretation of the field of
inertia, or the inerton field, surrounding the moving particle in
the range covered by the amplitude $\Lambda $ of the inerton
cloud. This means that inertons accompanying a moving particle
represent a substructure of the particle's matter waves.

\section*{5. The contraction of mass}

\hspace*{\parindent} The behavior of a moving canonical particle
pulling its deformation coat and surrounded by the cloud of inertons
can be studied also in the framework of a hydrodynamic approach. In
hydrodynamics, the notion of point particle is limited by the proper
size of the considered element of the liquid. This size is enormous
as compared with the particle size. The motion of such an element in
the space tessel-lattice treated as a world fluid can be described
by the basic equation of hydrodynamics
\begin{equation}
\label{eq23} \rho \frac{{d{\kern 1pt} v}}{{d{\kern 1pt} t}} = -
\nabla {\kern 1pt} P
\end{equation}

\noindent where $\rho$ is the liquid element density and
\textit{P} is the pressure of the liquid on the moving element. We
assume that the motion is adiabatic when the change in pressure on
the side of the liquid upon the element is proportional to the
variation in density of this element and then
\begin{equation}
\label{eq24} \left( {\frac{{\partial {\kern 1pt} P}}{{\partial
\rho} }} \right)_{\rm entropy} = c^{{\kern 0.6pt}2}
\end{equation}

\noindent
where \textit{c} is the maximum velocity for this liquid, i.e. sound
velocity.

Equation (23) is non-linear and consequently allows for multiple
solutions. However, there is only one possibility when this
equation becomes linear, and, therefore, there is a single
solution. That situation is realized when we examine the motion of
an element at the limits of tolerance, i.e. when space-time
derivatives become discrete. Note that recently Dubois [26,27] has
successfully applied space-time derivatives deduced from forward
and backward derivatives in computation for the study of quantum
relativistic systems and electromagnetism. By the way, by Dubois
[27] the masses of particles could be interpreted as properties of
space-time shifts, which is also in support of our definition of
mass as a local deformation of space.

The non-stationary motion of the element of our liquid allows the
representation of the equation of motion (23) in a discrete form
[16]. At the discrete consideration the substantial derivative has
to be transformed as follows
\begin{equation}
\label{eq25} \frac{{d{\kern 1pt}  f }}{{d{\kern 0.5pt}s}} =
\mathop {\lim}\limits_{\Delta s  {\kern 1pt}  \to  {\kern 1pt}
\Delta s_{\kern 0.5pt 0}} \frac{{f\left( {s + \Delta {\kern 1pt}
{\kern 1pt} {\kern 1pt} s} \right) - f\left( {s} \right)}}{{\Delta
{\kern 1pt} {\kern 1pt} s}} = \mathop {\lim}\limits_{\Delta s
{\kern 1pt} \to {\kern 1pt} \Delta s_{\kern 0.5pt 0}} \,{\kern
1pt} \frac{{\Delta {\kern 1pt} f}}{{\Delta {\kern 1pt} s}}
\end{equation}

\noindent where $\Delta {\kern 0.6pt} s_{0} $ stands for the size
of the element $\Delta {\kern 1pt}l_{0} $ or the time interval
$\Delta {\kern 1pt} t_{0} $ needed to pass the section equal to
the element's size. The submicroscopic consideration says that the
smallest size of the macroscopic element along the element path
has to be restricted by the perturbance of space caused by the
inerton cloud; therefore, $\Delta  l_{0} = \lambda /2$ and $\Delta
t_{0} = T/2$. Let $\upsilon $ and ${\kern 1pt} {\kern 1pt} {\kern
1pt} \rho _{0} $ be the initial values of the velocity and the
density of the element in question, respectively. Then Eqs. (23)
and (24) can be written as
\begin{equation}
\label{eq26} \rho {\kern 1pt} \frac{{\Delta  v} }{{T/2}} = -
{\kern 1pt} {\kern 1pt} {\kern 1pt} \frac{{\Delta  P}}{{\lambda
/2}},
\end{equation}
\begin{equation}
\label{eq27} {\kern 1pt} \frac{{\Delta {\kern 1pt} P}}{{\Delta
{\kern 1pt} \rho} } = c^{{\kern 0.6pt}2}.
\end{equation}

Eqs. (26) and (27) result in equation
\begin{equation}
\label{eq28} \rho {\kern 2pt} \Delta  v {\kern 1pt} {\kern 1pt}
\frac{{\lambda }}{{T}}\,\,\, = \,\, - c^{{\kern 0.6pt} 2}\Delta
\rho .
\end{equation}

In Eq. (28) $\lambda /T = v$ (see expression (20)). Since by
definition ${\kern 1pt} \Delta {\kern 1pt} f = f_{\rm current} -
f_{0} $ where $f_{0} $ stands for the initial value of the
function, we obtain $ \Delta v = - v $ for odd sections $\lambda
/2$ of the element path where the element velocity decreases to
zero. When the speed of the element changes from the maximum to
the minimum magnitude, the pressure, on the contrary, changes from
the minimum ($P$) to the maximum ($ P_{0} $) magnitude and
therefore ${\kern 1pt} \Delta P = P - P_{0} $. Then from Eq. (27)
we get ${\kern 1pt} \Delta  \rho = \rho - \rho _{0} $.
Substituting the corresponding values to equation (28) we arrive
at the expression
\begin{equation}
\label{eq29} \rho {\kern 1pt} {\kern 1pt} {\kern 1pt} =
\frac{{\rho _{0}} }{{{\kern 1pt} 1 - v^{2}/c^{2}}}.
\end{equation}

In the case of even sections $\lambda /2$ of the element path
where the element velocity increases from zero to $\upsilon $, we
get $\Delta v = v $ and then $ \Delta {\kern 1pt} \rho = \rho _{0}
- \rho $. These values of the parameters transform Eq. (28) to the
same expression (29).

Inasmuch as by definition (13), the mass is inversely proportional
to the volume, which it occupies in the tessel-lattice, we may write
for the mass of the liquid element ${\kern 1pt} {\kern 1pt} m_{0}
\propto 1/V_{0} $. Therefore, $\rho {\kern 1pt} _{0} {\kern 1pt}
{\kern 1pt} = m_{0} /V_{0} {\kern 1pt} {\kern 1pt} {\kern 1pt}
{\kern 1pt} {\kern 1pt} \propto {\kern 1pt} {\kern 1pt} {\kern 1pt}
{\kern 1pt} {\kern 1pt} \left( {1/V_{0}} \right)^{2}$, and we derive
from expression (29)
\begin{equation}
\label{eq30} m{\kern 1pt} {\kern 1pt} {\kern 1pt} = \frac{{m_{0}}
}{{\sqrt {1 - \upsilon ^{2}/c^{2}}} }.
\end{equation}

This result is in line with that (6) formally obtained by using
the fractality of particle-giving deformations.

Thus the model correctly describes the increase of mass and reduction of
dimensions of a moving object along a path and in this respect the model
complies with the Lorentz hypothesis on the change of the geometry of a
particle in motion.

\section*{6. Gravitational potential}

\hspace*{\parindent} What is a mechanism of the emission of
inertons from a moving particle? What is the reason to turn
emitted inertons back to the particle? What do inertons transfer
exactly?

To answer these questions we first of all should mention that a
gentle hint to the appropriate mechanism was given by de Broglie
[28,29]. De Broglie indicated that the corpuscle dynamics was the
basis for the wave mechanics. With the variational principle, he
obtained and studied the equations of motion of a massive point
reasoning from the typical Lagrangian (9). The study showed that
the dynamics had the characteristics of the dynamics of the
particles with a variable proper mass. Nevertheless, in de
Broglie's research the velocity $v$ of the point still was
constant along a path.

Owing to the availability of the crystallite surrounding the
particle in the tessel-lattice, we should examine the interaction of
the particle with the crystallite's oscillating mode. An appropriate
mechanism has been described in Ref. [24] in detail. The main
aspects of the particle - oscillating mode interaction are as
follows. At each collision of the moving particle with the mode,
whose energy is supported by the ambient tessel-lattice, the
particle loses a fragment $\delta {\kern 1pt} V$\textit{} of its
total deformation $V_{\rm particle} {\kern 2pt}\sqrt {1 - v
^{2}/c^{2}} $ (in line with the fractal decomposition principle
[11]) or, in other words, the inertial mass (30) decreases on a
value of the inerton mass $\mu _{1} $. The created quasi-particle,
i.e. inerton, is characterized by the kinetic energy and the
momentum obtained from the mode and the particle, which enables the
inerton to go off the particle.

Inertons are emitting from the particle until the particle is
moving. Since the value of the particle velocity oscillates along
the particle path, $v \to 0$ in each odd section $\lambda /2$ and $0
\to v$ in each even section $\lambda /2$, the decay of the total
mass of the particle (so-called the relativistic mass) (30) must to
follow the behavior of the particle, namely, the total particle mass
has also to oscillate along the particle's path: $m \to {\kern 1pt}
{\kern 1pt} 0$ and then $0 \to {\kern 1pt} {\kern 1pt} m$. The total
particle mass (3) comes to the inerton cloud that spreads over the
space tessel-lattice down to a distance $r = \Lambda $.

A number of inertons created at the decay of the particle mass in
each odd section $\lambda /2$ is huge and can be evaluated by the
number \textit{N} of collisions of the particle with
tessel-lattice's cells in this section, i.e. $N \sim  \lambda
/l_{\rm Planck} $ (recall $l_{\rm Planck}  \sim 10^{ - 35}$ m is the
size of the tessel-lattice's cell); for instance, if $\lambda = 0.1$
nm, the number of inertons created by the particle is on the order
of $N = \lambda /l_{\rm Planck}   \sim   10^{26}$. The value of the
mass $\mu _{1} $ of an inerton can widely varies and by the
assessment [29] $\mu _{1} {\kern 1pt} {\kern 1pt} {\kern 1pt}
\approx {\kern 1pt} {\kern 1pt} {\kern 1pt} 10^{ - 85}$ to $ 10^{ -
45}$ kg.

Emitted inertons should come back to the particle bringing fragments
of its deformations in even section $\lambda /2$ of the particle
path. In other words, absorbed inertons should restore the initial
state of the moving particle, i.e. its mass and velocity. The
simplest mechanism of such behaviour of inertons, which strings them
to the particle, can be understood only in the framework of a
\textit{general dynamics of the tessel-lattice}.

Having returned back to the particle, the inerton in question has
initially to come to a stop. The resetting of the inerton means that
this the elastic tessel-lattice that guides it at a distance
$\Lambda $ from the particle and then replaces again. Therefore, the
emitted inerton should undergo an elastic influence on the side of
the tessel-lattice. Hence we have to supplement the Lagrangian (14)
by additional terms that will lead to the Euler-Lagrange equations
describing changes in both an elastic property of the inerton and
that of the tessel-lattice. What can change in the bound inerton?
Obviously the degree of its deformation can only change, i.e. the
value of the inerton mass. In other words, this local deformation
migrating from cell to cell slowly drops while inducing a
\textit{rugosity} in ongoing cells, which can also be called the
\textit{tension of the tessel-lattice} in physics terms, if applied
to the cloud of inertons as a whole. The induction of the rugosity,
or tension, does not destroy the morphism of cells, but translates
the cells from their equilibrium positions in the tessel-lattice.
The tension is removed by the energy stored in the tessel-lattice,
or in other words, the tessel-lattice restores its initial state in
which every cell occupies its own equilibrium position.

Thus the Lagrangian (14) is extended to
\begin{equation}
\label{eq31}
\begin{array}{l}
 L = - m_{0} c^{2}\left\{ {1 - \frac{{1}}{{m_{0} c^{2}}}\left[ {m_{0} \dot
{x}^{2} - \frac{{2\pi} }{{T}}\sqrt {m_{0} \mu _{0}}  {\kern 1pt}
\left( {x\dot {\chi}  + v {\kern 1pt} \chi}  \right) + \mu _{0}
\dot {\chi
}^{2}} \right]} \right. \\
 \quad \quad \quad \;\;\,\quad \,\left. {\,\,\,\, + \left[
{\frac{{T^{2}}}{{2{\kern 0.6pt} m_{0}^{2}} } {\kern 1pt} \dot
{m}^{2}\,\,\, + \,\,\,\frac{{T^{2}}}{{2\Lambda ^{2}}}\,\,\dot
{\vec {\xi} }^{{\kern 2pt} 2}\,\,\, - \,\,\,\frac{{T}} {{m_{0}}
}\,\,\dot {m} \, \nabla {\kern 0.6pt} \vec {\xi} }
{\kern 2pt}\right]\ } \right\}^{1/2}. \\
 \end{array}
\end{equation}

Here $m = m{\kern 1pt} {\kern 1pt} {\kern 1pt} \left( {\vec
{r},\,\,t} \right)$ is the current mass of the \{particle + inerton
cloud\}-system; $\vec {\xi}  = \vec {\xi} {\kern 1pt} {\kern 1pt}
\left( {\vec {r},\,\,t} \right)$ is the current value of the
rugosity of the tessel-lattice in the range covered by the system.
Geometrically the rugosity $\vec {\xi} $ depicts the state in which
the tessel-lattice cells in the region covered by the inerton cloud
do not have any volumetric deformation (i.e. the cells become
massless), but are slightly shifted from their equilibrium positions
in the tessel-lattice along the particle vector velocity $\vec {v}
$.

Proceeding to the Euler-Lagrange equations for variables $m$ and $\vec {\xi
},$ we have to use them in the form (due to the term $\nabla {\kern 1pt}
{\kern 1pt} \vec {\xi} $, see e.g. Ref. [30])
\begin{equation}
\label{eq32} \frac{{\partial} }{{\partial {\kern 1pt} {\kern 1pt}
t}}\frac{{\partial L}}{{\partial {\kern 1pt} \dot {q}}} -
\frac{{\delta {\kern 1pt} L}}{{\delta {\kern 1pt} q}} = 0
\end{equation}

\noindent
where the functional derivative
\begin{equation}
\label{eq33} \frac{{\delta {\kern 1pt} L}}{{\delta
{\kern 1pt} q}} = \frac{{\partial L}}{{\partial {\kern 1pt} q}} -
\frac{{\partial} }{{\partial {\kern 1pt} x}}\frac{{\partial
L}}{{\partial {\kern 1pt} {\kern 1pt} \left( {\partial {\kern 1pt}
q/\partial {\kern 1pt} x} \right)}} - \frac{{\partial }}{{\partial
{\kern 1pt} y}}\frac{{\partial L}}{{\partial {\kern 1pt} {\kern
1pt} \left( {\partial {\kern 1pt} q/\partial {\kern 1pt} y}
\right)}} - \frac{{\partial} }{{\partial {\kern 1pt}
z}}\frac{{\partial L}}{{\partial {\kern 1pt} {\kern 1pt} \left(
{\partial {\kern 1pt} q/\partial {\kern 1pt} z} \right)}}.
\end{equation}

The equations for $m$ and $\vec {\xi} $ obtained from Eqs. (32)
and (33) are the following
\begin{equation}
\label{eq34} \frac{{\partial ^{{\kern 1pt} 2}m}}{{\partial {\kern
1pt} {\kern 1pt} {\kern 1pt} t^{2}}}{\kern 1pt} {\kern 1pt} {\kern
1pt} - {\kern 1pt} {\kern 1pt} {\kern 1pt} \frac{{m_{0}}
}{{T}}\nabla \dot {\vec {\xi} } = 0,
\end{equation}
\begin{equation}
\label{eq35} \frac{{\partial ^{{\kern 1pt} {\kern 1pt} 2}{\kern
1pt} \vec {\xi }}}{{\partial {\kern 1pt} {\kern 1pt} t^{2}}}{\kern
1pt} {\kern 1pt} {\kern 1pt} - {\kern 1pt} {\kern 1pt} {\kern 1pt}
\frac{{\Lambda ^{2}}}{{m_{0} T}}{\kern 1pt} {\kern 1pt} {\kern
1pt} \nabla {\kern 1pt} \dot {m} = 0.
\end{equation}

To solve equations (34) and (35), we have to set initial
conditions to variables $m{\kern 1pt} {\kern 1pt} {\kern 1pt}
\left( {\vec {r},{\kern 1pt} {\kern 1pt} {\kern 1pt} {\kern 1pt}
t} \right)$ and $\vec {\xi} {\kern 1pt} {\kern 1pt} \left( {\vec
{r},{\kern 3pt}  t} \right)$. It is obvious that the initial
conditions are
\begin{equation}
\label{eq36} m{\kern 1pt} {\kern 1pt} \left( {\vec {r},{\kern 1pt}
{\kern 1pt} {\kern 1pt} {\kern 1pt} 0} \right) = m{\kern 1pt}
{\kern 1pt} \left( {\vec {r}} \right), \quad \left.
{\frac{{\partial {\kern 1pt} {\kern 1pt} {\kern 1pt} m}}{{\partial
{\kern 1pt} {\kern 1pt} {\kern 1pt} t}}} \right|_{t = 0} = 0;
\end{equation}
\begin{equation}
\label{eq37} \vec {\xi} {\kern 1pt} {\kern 1pt} \left( {r,{\kern
1pt} {\kern 1pt} {\kern 1pt} {\kern 1pt} 0} \right) = \vec {\xi}
{\kern 1pt} {\kern 1pt} \left( {r} \right), \quad \left.
{\frac{{\partial {\kern 1pt} {\kern 1pt} \vec {\xi} }}{{\partial
{\kern 1pt} {\kern 1pt} t}}} \right|_{t = 0} = 0.
\end{equation}

The boundary conditions are
\begin{equation}
\label{eq38} m{\kern 1pt} ( {{\kern 1pt} l_{\rm Planck},{\kern
1pt} t}) = m (t), \ \    m ( \Lambda ,{\kern 1pt} t ) = 0, \ \
 {\frac{{\partial {\kern 1pt} m}}{{\partial {\kern 1pt}\vec
{r}}}} \Big|_{r = l_{\rm Planck}} = 0, \ \  {\frac{{\partial
m}}{{\partial {\kern 1pt} \vec {r}}}} \Big|_{r = \Lambda} = f(t);
\end{equation}
\begin{equation}
\label{eq39} \vec {\xi} ( l_{{\kern 0.5pt}\rm Planck},  {\kern
1pt} t ) = 0, \ \ \vec {\xi} {\kern 1pt} {\kern 1pt} (\Lambda
,{\kern 1pt} t ) = \vec {\xi} (t), \ \ {\frac{{\partial {\kern
1pt} \vec {\xi }}}{{\partial {\kern 1pt} \vec {r}}}} \Big|_{r =
l_{\rm Planck}} = F (t), \ \     {\frac{{\partial {\kern 1pt}
\mathord{\buildrel{\lower3pt\hbox{$\scriptscriptstyle\rightharpoonup$}}\over
{\xi} }} }{{\partial {\kern 1pt} \vec {r}}}} \Big|_{r = \Lambda} =
0.
\end{equation}

The conditions (36) to (39) show that the mass is initially has
located in the center of coordinates of the system studied, i.e. in
the particle, such that $m{\kern 1pt} \left( {\vec {r},{\kern 1pt}
t} \right)|{\kern 1pt} _{l_{{\kern 0.5pt} \rm Planck}, {\kern 1pt}
{\kern 1pt} 0}  =  m_{0} {\kern 0.5pt} /\sqrt {1 - v ^{2}/c^{2}}$.
Besides, these conditions mean that the behavior of the mass of the
\{particle + inerton cloud\}-system and that of the rugosity of the
tessel-lattice are opposite in phase.

The conditions (36) to (39) allow one to transform equations (34)
and (35) to the form ($\Delta $ is the Laplace operator)
\begin{equation}
\label{eq40} \frac{{\partial ^{{\kern 1pt} 2}m}}{{\partial {\kern
1pt} {\kern 1pt} {\kern 1pt} t^{2}}}{\kern 1pt} {\kern 1pt} {\kern
1pt} - {\kern 1pt} {\kern 1pt} {\kern 1pt} c^{2}\,\Delta {\kern
1pt} {\kern 1pt} {\kern 1pt} m = 0,
\end{equation}
\begin{equation}
\label{eq41} \frac{{\partial ^{{\kern 1pt} {\kern 1pt} 2}{\kern
1pt} \vec {\xi }}}{{\partial {\kern 1pt} {\kern 1pt} t^{2}}}{\kern
1pt} {\kern 1pt} {\kern 1pt} - {\kern 1pt} {\kern 1pt} {\kern 1pt}
\frac{{\Lambda ^{2}}}{{m_{0} T}}{\kern 1pt} {\kern 1pt} {\kern
1pt} \nabla {\kern 1pt} \dot {m} = 0.
\end{equation}

Since the system studied features the radial symmetry, variables
$m$ and $\vec {\xi} $ are functions of only the distance $r$ from
the particle and the proper time \textit{t} of the \{particle +
inerton cloud\}-system. In this case we preserve only radial
components in the both variables, which enables to rewrite
equations (40) and (41) in the spherical coordinates as follows
\begin{equation}
\label{eq42} \frac{{\partial ^{{\kern 1pt} 2}m}}{{\partial {\kern
1pt} {\kern 1pt} {\kern 1pt} t^{2}}} -
c^{2}\frac{{1}}{{r}}\frac{{\partial ^{{\kern 1pt} 2}\left(
{r{\kern 1pt} {\kern 1pt} m} \right)}}{{\partial {\kern 1pt}
{\kern 1pt} r^{2}}} = 0,
\end{equation}
\begin{equation}
\label{eq43} \frac{{\partial ^{{\kern 1pt} 2}\xi} }{{\partial
{\kern 1pt} {\kern 1pt} {\kern 1pt} t^{2}}}\,\, - {\kern 1pt}
{\kern 1pt} {\kern 1pt} \frac{{\Lambda ^{2}}}{{m_{0} T}}{\kern
1pt} {\kern 1pt} {\kern 1pt} \frac{{\partial {\kern 1pt} {\kern
1pt}} }{{\partial {\kern 1pt} {\kern 1pt} r}}\frac{{\partial
{\kern 1pt} {\kern 1pt} m}}{{\partial {\kern 1pt} {\kern 1pt} t}}
= 0
\end{equation}

\noindent where the Laplace operator $\Delta $ is presented in the
spherical coordinates as $\Delta {\kern 1pt} {\kern 1pt} m =
{\textstyle{{1} \over {r}}}{\textstyle{{\partial ^{2}} \over
{\partial {\kern 1pt} r^{2}}}}\left( {r{\kern 1pt} m} \right)$.

Thus the availability of the radial symmetry and the conditions
(36) to (39) allow the solutions to equations (42) and (43) in the
form of standing spherical waves, which exhibit the dependence
$1/r,$
\begin{equation}
\label{eq44} m{\kern 1pt} {\kern 1pt} {\kern 1pt} \left(
{r,\,\,\,t} \right) = C_{1} {\kern 1pt} \frac{{m_{0}} }{{r}}{\kern
1pt} {\kern 1pt} {\kern 1pt} \cos \left( {\frac{{\pi {\kern 1pt}
r}}{{2\Lambda} }} \right){\kern 1pt} \left| {\kern 1pt} {\cos
\left( {\frac{{\pi {\kern 1pt} t}}{{2{\kern 1pt} {\kern 1pt} T}}}
\right)} \right|,
\end{equation}
\begin{equation}
\label{eq45} \xi {\kern 1pt} {\kern 1pt} \left( {r,\,\,\,t}
\right) = C_{2} {\kern 1pt}\frac{{\xi _{0}} }{{r}}{\kern 1pt}
{\kern 1pt} {\kern 1pt} {\kern 1pt} \sin \left( {\frac{{\pi {\kern
1pt} r}}{{2\Lambda} }} \right) \,  \left( { - 1} \right)^{\left[
{t/2T} \right]}{\kern 1pt} \sin \left( {\frac{{\pi {\kern 1pt}
t}}{{2{\kern 1pt} {\kern 1pt} T}}} \right)
\end{equation}

\noindent where we omit the radical $\sqrt {1 - v ^{2}/c^{2}} $ at
the mass $m_{{\kern 0.5pt}0}{\kern 0.6pt}$ and introduce the
notation $\left[ {t/2T} \right]$ meaning an integral part of the
integer $t/2T$. The dimensionality of integration constants
$C_{1,2} $ corresponds to length and one can put $C_1 = l_{\rm
Planck} \approx 10^{ - 35}$ m and $C_{{\kern 0.5pt}2} = \Lambda$.

The solution (44) shows that the amplitude of mass of the inerton
cloud
\begin{equation}
\label{eq46} \frac{m_0 }{r}{\kern 1pt}  \cos \left( {\frac{{\pi
{\kern 1pt} r}}{{2\Lambda} }} \right),
\end{equation}

\noindent represents the particle mass distribution in the range
from $r = l_{\rm Planck} $ to $r = \Lambda $ where $\Lambda =
\lambda {\kern 1pt} {\kern 1pt} {\kern 1pt} c/v $ is the amplitude
of the particle's inerton cloud.

The solution (45) depicts a similar pattern for the behavior of the
rugosity, or tension, of the tessel-lattice in ambient space.

In the region of space $l_{{\kern 0.6pt}\rm Planck}  \ll r {\kern
1pt} \ll {\kern 1pt} \Lambda $, the time-averaged distribution of
the \{particle + inerton cloud\}-mass becomes
\begin{equation}
\label{eq47} m\,\,\left( {r} \right)\, \cong \, l_{{\kern
0.6pt}\rm Planck} \, \frac{{m_{{\kern 1pt} 0}} }{{r}}.
\end{equation}

In this region of space, the tension of space, as follows from
expression (45), $\xi {\kern 1pt} {\kern 1pt} \left( {r} \right)
\cong 0$.

Thus in mathematics terms, one can say that a local fractal
deformation of the space tessel-lattice when moves induces the
rugosity in the tessel-lattice. When the local deformation is
distributed in space, it forms a deformation potential $ \propto
\,1/r$ that spreads down to a distance ${\kern 1pt} r = \Lambda $
from the core-cell, i.e. particle. \textit{In the range covered by
the deformation potential cells of the tessel-lattice are found in
the contraction state and it is this state of space, which is
responsible for the phenomenon of the gravitational attraction.}

In physics terms, the total mass (30) of the particle periodically
decays (in the section equals the de Broglie wavelength $\lambda
$) transferring to the inertons cloud that induces the mass field,
i.e. deformation potential, $\propto \, 1/r$ (47), in the range up
to a distance $r < \Lambda $, which is identical to the
gravitational potential of the particle.

Does the distribution of mass field (47) of an isolated particle
lead to Newton's potential of a macroscopic object? Yes, it does.
A macroscopic object is a many-particle system in which particles
(atoms is a solid or protons in a star) are tied by means of their
inerton clouds in a uniform system. Such unification gives rise to
coherent collective vibrations of atoms/protons (acoustic
vibrations). As has been argued early [31], those are amplitudes
$A$ of vibrating atoms that in a condensed matter play the role of
the de Broglie wavelength $\lambda $ of free particles. Therefore
in a condensed matter relation (21) is replaced by
\begin{equation}
\label{eq48} \Lambda = A {\kern 1pt} {\kern 1pt} c/v
\end{equation}

\noindent where the velocity $v $ should be the same for every
atom, because this is the sound velocity (the group velocity of
vibrating atoms in acoustic modes). The amplitude $A$ depends on
the number of atoms involved in a vibrating mode, i.e. $A_{{\kern
1pt} s} = 2A_{1} \cdot s/\aleph $ where $s =
2,\,\,3,\,...,\,\,\left( {\aleph - 1} \right)$ is the number of
atoms in the \textit{s}th mode and $\aleph $ is the total number
of object's atoms. In such a way $A_{{\kern 1pt} s} $ should be
regarded as the wavelength of \textit{s}th mode/harmonic of the
object studied. The longest mode corresponds to collective
vibrations of all atoms of the object, which means that all
$\aleph $ atoms generate the total collective inerton cloud with
the wavelength $A_{{\kern 1pt} \aleph}  $ in the ambient space.
Thus expression (47) holds also for the rest mass of a macroscopic
object.

Multiplying both parts of expression (47) by a factor $ - G/l_{\rm
Planck}$ where $G$ is the Newton constant of gravitation, we
obtain the so-called potential gravitational energy, or Newton's
gravitational potential, of the object in question
\begin{equation}
\label{eq49} U\left( {r} \right)\, = \, - {\kern 1pt}
G\,\frac{{m_{0}} }{{r}}
\end{equation}

\noindent where the Newton constant simple plays the role of a
dimensional coefficient. As provided by relation (48), the
potential (49) spreads from the object down to a huge distance $r
< \Lambda _{{\kern 0.6pt}\aleph} {\kern 2pt} c/v \sim \left( {{\rm
V}_{\rm object} /{\rm v}_{\rm particle}} \right){\kern 1pt} {\kern
1pt} {\kern 1pt} c/v $, where $V_{\rm object} $ and $\rm v_{\rm
particle} $ are typical volumes of the object and the object's
effective particle, respectively.

\section*{7. Experimental verification}

\hspace*{\parindent} To check the model of gravitation described
above one should test excitations of the space tessel-lattice, which
is considered in the given model as a primary substrate. Let us
dwell on major experimental results available at present.

\medskip

\textit{7.1. A resonator of inertons.} The impact of inerton waves
on the behavior of atoms in metals was studied theoretically and
then observed experimentally as changes in the fine morphological
structure of specimens by the high-resolution electron-scanning
microscope in paper [32]. In the experiment we used a resonator of
inerton waves of the Earth, which represented a small projective
model of the terrestrial globe. In essence the paper is the first
reliable demonstration of the existence of the so-called aether wind
generated by the Earth at its motion through the world aether, or
space that is most appropriate for our consideration.

\bigskip

\textit{7.2. Anomalous photo-electric effect.} In paper [33] electrons
moving in atoms were treated as entities surrounded by their inerton clouds.
The investigation of the interaction between such entities and a photon flux
was curried out in detail. The major peculiarity of the theory proposed was
the effective cross-section of electrons significantly enlarged due to their
inerton clouds spread around the electrons.

It was shown that a number of different experiments aimed at the
study of laser-induced gas ionization were in agreement with the
theoretical results prescribed by the inerton theory, namely: (a)
the concentration of ionized atoms was directly proportional to
the peak laser pulse intensity and the time to the second power;
(b) the prediction of temporal dependence of the threshold
intensity was in accord with the experimental data; (c) the
experimental results on the number of ions created by the laser
pulse as a function of the pulse intensity checked well with the
description in terms of the anomalous photo-electric effect
constructed; (d) the breakdown intensity threshold measured as a
function of pressure or gas density fitted well with the theory;
(e) an anomalous value of electrons released from atoms of gas at
high energies of the laser beam correlated with the inerton
theory.

Electron emission from a laser-irradiated metal was also correctly described
by the inerton theory, namely, the theory was in consistent with (a) the
experimental results indicated that the photo-electric current was a linear
with light intensity; (b) the data showing that the maximum energy of the
emitted electron is a function of light intensity.

\bigskip

\textit{7.3. Oscillations of hydrogen atoms caused by the inerton
field.} The inerton concept was justified [31] at the employing
for the experiment on the study of fine dynamics of hydrogen atoms
in the $\rm KIO_{3} \cdot HIO_{3} \,$crystal whose FT IR spectra
in the 400 to 4000 cm$^{-1}$ spectral range showed unexplainable
sub maximums. Features observed in the spectra were unambiguously
interpreted [31] as caused by the hydrogen atoms' matter waves
that induced the mean inerton field contributing to the paired
potential of hydrogen-hydrogen interaction. This admitted solution
in the cluster form. The number of hydrogen atoms entered a
cluster was assessed and its spectrum was calculated. It was
inferred [31] that those were sub maximums in the total spectrum
of the crystal, which emerged due to the collective oscillations
of hydrogen atoms interacting in clusters through the inerton
field.

\bigskip

\textit{7.4. A deviation from the inverse square law for gravity.} This
theme attracts a wide attention of those researchers who try to test gravity
in a microscopic range. It is assumed [34] that a new kind of interaction
coexists with the conventional gravity, which has to modify Newton's
gravitational law. The non-Newtonian potential is chosen in the form of the
Yukawa potential in such a way that the modified Newtonian potential for
masses $m_{i} $ and $m_{j} $ is written as follows

\begin{equation}
\label{eq50} U\left( {r} \right) = - {\kern 1pt} {\kern 1pt}
G{\kern 1pt} {\kern 1pt} {\kern 1pt} {\kern 1pt} \frac{{m_{i}
{\kern 1pt} m_{j}} }{{r}}{\kern 1pt} {\kern 1pt} {\kern 1pt}
{\kern 1pt} \left( {\,1 + \alpha \,e^{ - {\kern 1pt} {\kern 1pt}
{\kern 1pt} r/Y}{\kern 1pt}}  \right)
\end{equation}

\noindent where $Y$ is the Yukawa distance over which the
corresponding force acts and $\alpha $ is a strength factor in
comparison with Newtonian gravity.

In the recent experiment [35] ultra-cold neutrons falling in the Earth
gravitational field are reflected and trapped in a cavity above a horizontal
mirror. It was revealed that the population of the ground state and the
lowest neutron states followed the quantum mechanical prediction (the
higher, unwanted neutron states were removing by an efficient absorber).
This very interesting precise experiment showed the availability of quantum
states for Newtonian gravity on the micrometer scale, which allowed the
authors to place limits for an additional gravity-like force in the range
from 1 to 10 µm with an explicit quantum tail down to 70 µm.

One can ask what is the origin of the correction introduced in
expression (50) to the classical Newton's law? Let us try to
answer the question drawing inertons.

Let in expression (50) $m_{i} $ stands for the Earth mass $m_{\rm
Earth} $ and $m_{j} $ stands for the neutron mass $m_{\rm n} $. In
the zero approximation, the Earth gravitational field should
attract a test point particle, i.e. the neutron, in accordance
with the classical Newton's gravitational law,
\begin{equation}
\label{eq51} U\left( {r} \right) = - {\kern 1pt} {\kern 1pt}
G{\kern 1pt} {\kern 1pt} {\kern 1pt} {\kern 1pt} \frac{{m_{\rm
Earth} {\kern 1pt} m_{\rm n}} }{{r}}
\end{equation}

However, in the next approximation we should take into account
that the neutron is not a point-like particle. In fact, in the
experiment described in Ref. [35] the neutron velocity was around
several meters per second such that the de Broglie wavelength
$\lambda $ varied from 40 nm to 100 nm. In conformity with
relationship (21) the neutron's inerton cloud spreads in
directions transversal to the neutron path down to $\Lambda =
\lambda {\kern 1pt} {\kern 1pt} c/v \sim  1$ m and this is the
real gravitational radius of the neuron. This means that the
neutron moving along its path falls via its inerton cloud within
the exchange gravitational interaction with the environment. Let
$M$ be the effective mass of the nearest environmental matter that
surrounds the moving neutron and $\mu $ be the mass transferable
by the neutron's inerton cloud. Then the following "kinetic"
equations can be proposed
\begin{equation}
\label{eq52} \frac{{d{\kern 0.3pt}\mu} }{{d {\kern 0.6pt}r}} =
a\mu - bM,
\end{equation}
\begin{equation}
\label{eq53} \frac{{d{\kern 0.3pt}M}}{{d{\kern 0.6pt}r}} = kM -
a\mu.
\end{equation}
Here in Eq. (52) the terms $a\mu $ and $bM$ respective describe
the decay of the inerton cloud's mass of a moving neutron and the
restoration of the inerton cloud's mass due to the mass input on
the side of the environment; in Eq. (53) the terms $kM$ and $a\mu
$ describe the outflow of the mass from the environment and the
input mass to the environment from the neutron's inerton cloud,
respectively. The variable $r$ in Eqs. (52) and (53) characterizes
the distance of the moving neutron to the detector (the Earth
surface). Equations (52) and (53) result in the following equation
for the inerton cloud's mass
\begin{equation}
\label{eq54} \frac{{d^{2}\mu} }{{dr^{2}}}\,\, - \,\,\,\left( {a +
k} \right){\kern 1pt} {\kern 1pt} \frac{{d\mu} }{{dr}}{\kern 1pt}
{\kern 1pt} {\kern 1pt} {\kern 1pt} - {\kern 1pt} {\kern 1pt}
{\kern 1pt} {\kern 1pt} a\left( {k - b} \right)\mu \,\, = \,0.
\end{equation}
Eq. (\ref{eq51}) has the solution
\begin{equation}
\label{eq55} \mu \left( {r} \right) = \mu _{\rm eff} \exp \left[ {
- \left( {\frac{{a + k}}{{2}} + \sqrt {\left( {\frac{{a +
k}}{{2}}} \right)^{2} + a{\kern 1pt} {\kern 1pt} \left( {k - b}
\right)}} \right)\,\,r} \right]
\end{equation}

\noindent where, as clear from Eqs. (52) and (53), $k - b
> 0$. Expression (55) can be rewritten in the contracted
form,
\begin{equation}
\label{eq56} \mu \left( {r} \right) = \mu _{\rm eff} {\kern 1pt}
\exp\left( { - Y{\kern 1pt} r} \right)
\end{equation}

\noindent and then the mass $m_{\rm n} $ of the neutron in
expression (51) should be added by the correction (56), i.e.
\begin{equation}
\label{eq57} m_{\rm n} \to \,\, m_{\rm n}  + \mu _{\rm eff} {\kern
1pt}\exp\left( { - {\kern 1pt} {\kern 1pt} {\kern 1pt} Y r}
\right),
\end{equation}

\noindent which immediately results in the modified Newton's law
(50). In the experiment [35] the parameter ${\kern 1pt} {\kern
1pt} Y$ was estimated as equal to several micrometers. Note that
the parameters entered expression (55) allow the evaluation in
principle.

Consequently, the origin of the Yukawa potential in expression
(50) is originated from the exchange of mass of the inerton cloud
of a falling particle with a matter surrounding the particle.

\bigskip

\textit{7.5. Circumstantial confirmations.} In paper [36] the
authors claimed that they observed $\psi$-wave functions of
electrons on metal surfaces. This automatically means that in fact
the $\psi$-wave function is not an abstract mathematical
construction, but a function that characterizes the real
perturbation of space surrounding the electron in the metal, i.e.
it specifies the inerton cloud of the electron.

Podkletnov [37,38] has made a demonstration of the loss of a part of the
mass of electrons in the superconductor state at special conditions, which
can serve as a good example supporting our theory of quantum gravity that
emerges due to the inerton interaction between quantum entities.

Changes in the structure of metal samples irradiated by a field of
undetermined nature (so-called the \textit{torsion field}) are displayed in
book [39] (the results by V. Maiboroda, A. Akimov et al.). The torsion
radiation was introduced pure theoretically by Shipov [39] as a primary
field that allegedly was dominating over a vague physical vacuum long before
its creation. Nevertheless, the named authors experimenting with metals have
used as a source of this new field concrete generators, which allows us to
suggest that they have dealt with the sources of the inerton waves.

The same as the author, Baurov [40] sees all particles as being conceived
from a unique corpuscle; moreover, some direct experimental evidences of the
interaction of matter with a sub quantum medium has indeed been demonstrated
by him.

Experimentalists who investigate low energy nuclear reactions claim that
they fix a new ``strange" radiation at the transmutation of nuclei, Urutskoev
et al. [41]; clearly that ``strange" radiation is nothing else as the inerton
field that is radiated from the appropriate inerton clouds of nucleons when
the latter are rearranging in nuclei. Gauging the radiation of an unknown
field from a rotating ferrite disc was performed by Benford [42]; the work
makes a demonstration of the transmutation of chemical entities under this
unknown radiation.

Egyptian pyramids are a remarkable example of an aether wind
trapping site on the Earth. The pyramids are still greatest
speechless puzzles for modern science. Their geometry and
orientation by the world sides point out that the pyramids were
built as resonators of inerton waves of the Earth. The prehistoric
Egyptian civilization was aware of the subtle properties of matter
and used the Earth inerton field in the pyramids for applied
purposes (note that the Egyptian pyramids were built over than
10,000 years ago, which already was proved scientifically); see
e.g. Ref. [43].

\bigskip

\textit{7.6. Further experimental protocols.} Former experiments
by several researchers [44,45] have confronted the recoding of
informational signals from stars other than electromagnetic ones,
which is known as Kozyrev's effect. We have proposed [46] a series
of protocols for testing the inerton radiation from stars and
planets; the measurements will be conducted by means of special
pyroelectric sensors constructed for this purpose, which will be
embedded in the focal volume of the telescope. The first
encouraging measurement of the inerton radiation was in fact
obtained along the West-East line at 20 Hz, which was associated
with proper rotation of the globe; besides, inerton signals at
frequencies 18 to 22 kHz, which came from the northern sky in a
universal time interval from 3 p.m. to 5 p.m., were recorded from
September 2004 to February 2005.

We [47] have just examined the phenomenon of radiation of physical
fields generated by the so-called Teslar watch on organic matter.
The Teslar watch contains a special chip that generates an unknown
field called the ``scalar Teslar wave"; the Teslar watch's radiation
very positively influences the human organism preserving it from the
environmental low frequency electromagnetic pollution, supports good
health, facilitates a fast relaxation of the nervous system, and so
on. Our study in fact has justified that the Teslar watch generates
nor the electromagnetic, neither ultrasound radiation. This is
nothing as the inerton radiation generated by a special electric
circuit, or the Teslar chip, embedded in the watch (two superimposed
electromagnetic waves whose amplitudes are shifted to 180$^{{\kern
0.5pt}\circ}$ are canceled, but an inerton flow that continues to
transfer the energy remains. This is quite possible, as it follows
from the study of the phenomenon of the electric charge and
principles of its motion in the tessel-lattice [47]). Five different
experiments were performed [48].

At last the inerton field is the fundamental one in nuclear
physics; it is this field that bounds nucleuses in nuclei
accounting for the reasons for the nuclear forces playing a role
of a control field in nuclear physics. Because of that, the
reconsideration of basic principles of nuclear physics has allowed
for a radically new approach to complete clean nuclear energy
[49].

\section*{8. Concluding remarks}

\hspace*{\parindent} In the present paper based on the rigorous
mathematical approach initiated by Michel Bounias we have disclosed
the founding principles of the constitution of the physical space
and shed light on the foundation of space-time. We have introduced
the determination of physical space as the fractal tessel-lattice
that is characterized by a primary element (a topological ball, or a
cell, or a superparticle) and defined the notion of mass and hence a
particle as a local deformation of space. The interaction of a
moving particle-like deformation with the surrounding tessel-lattice
involves a fractal decomposition process that supports the existence
and properties of previously introduced inerton clouds as associated
to particles.

The further study has allowed one to account for the inertial and
the gravitational masses of a particle, which in fact in line with
de Haas' [21] research aimed at the unification of Mie's theory of
gravitation and de Broglie's wave mechanics. It has been shown that
the so-called relativistic mass (30) is responsible for the
induction of the gravitational potential of a moving canonical
particle. Namely, the total particle mass (30) is allocated in the
particle's inerton cloud transferring an additional reduction to
surrounding cells of the tessellation space. The particle's inertons
spread to the range $\Lambda \approx \lambda \,c/v $ from the
particle where $v $ and $\lambda $ respective are the particle's
velocity and de Broglie wavelength. Therefore, a canonical particle
with a mass $m$ being moving generates the mass field, or the
gravitational mass, that obeys the law (47), and this consideration
remains valid in the case of a classical object.

The particle's inertons migrate from cell to cell by relay-race
mechanism and spread to the distance $\Lambda $ as a typical
advancing spherical wave. Since any spherical wave is characterised
by the inverse law $1/r,$ this immediately results in the
distribution of the inerton field proportional to $1/r$ around the
particle, or a classical object. This is the inner reason for
Newton's gravitational law. Thus the phenomenon of gravitational
attraction becomes complete clear: inertons carry fragments of the
inertial deformation, i.e. mass, of the object to the surrounding
space forming a deformation potential, i.e. the mass field, in which
each cell of the tessel-lattice undergoes an additional contraction
in accord with the rule (47).

The quantization of the gravitation is caused by the periodical
process of emission and absorption of inerton clouds by a moving
particle. This process intrudes itself upon the space in which the
motion occurs. Any cyclic motion allows the description in terms
of the Hamilton-Jacobi formalism. In our case the minimum of
increment of the action becomes equals to $\hbar$, which means
that it is the space that exerts control over the behavior of the
particle [15,16]. Hence submicroscopic mechanics is in fact
quantum and moreover is readily transformed to conventional
quantum mechanics developed in an abstract (phase, Hilbert, and
etc.) space.

The submicroscopic concept forbids the existence of so-called
gravitational waves of general relativity [32]. Loinger [50] has
proved the same, though he has started from the conventional
Gilbert-Einstein equations; namely, he has shown that Einstein's
gravitational waves are not a realistic solution.

The submicroscopic concept stated above has successfully been verified
experimentally, but rather in microscopic and intermediate ranges. The
question arises whether the theory can be adjusted with general relativity
that works in a macroscopic range. Probably the submicroscopic concept can
be able to account for those crucial experiments that have been formally
predicted by general relativity, namely: the motion of the perihelion of a
planet, the deviation of light by a star, and the red shift of light under
the gravitation. These studies will be conducted in the nearest future.

\end{document}